\documentclass[a4paper,fleqn]{cas-dc}
\usepackage[numbers]{natbib}
\usepackage{hyperref}
\usepackage{url}
\usepackage{xcolor}
\usepackage{latexsym}
\usepackage{framed,multirow}
\usepackage{amsmath,amssymb,amsfonts}
\usepackage{algorithmic}
\usepackage{graphicx}
\usepackage{subfig}
\usepackage{textcomp}
\usepackage{lineno}
\usepackage{multirow}
\usepackage{scalerel}
\usepackage{mathrsfs}
\usepackage{booktabs} 
\usepackage{rotating}
\usepackage{tikz}
\newcommand{\cor}{\textcolor{red}}
\begin{document}
\let\WriteBookmarks\relaxe
\def\floatpagepagefraction{1}
\def\textpagefraction{.001}
\shorttitle{Unsupervised seq2seq learning for automatic SQ assessment in multi-channel impedance-based hemodynamic monitoring}
\shortauthors{Hyun and Kim et~al.}
\title[mode = title]{Unsupervised sequence-to-sequence learning for automatic signal quality assessment in multi-channel electrical impedance-based hemodynamic monitoring}
\cortext[cor1]{Equal contribution: C.M Hyun and T.-G. Kim.}
\cortext[cor2]{Corresponding authors: C.M. Hyun and K. Lee.}
\author[1]{Chang Min Hyun}[type=editor,orcid=0000-0002-7072-7489]
\ead{chammyhyun@yonsei.ac.kr}
\cormark[1]
\author[2]{Tae-Geun Kim}[]
\cormark[1]
\author[3]{Kyounghun Lee}[orcid=0000-0002-8520-8999]
\ead{imlkh84@gmail.com}
\address[1]{School of Mathematics and Computing (Computational Science and Engineering), Yonsei University, Seoul, Republic of Korea}
\address[2]{Department of Physics, Yonsei University, Seoul, Republic of Korea}
\address[3]{Department of Biomedical Engineering, School of Medicine, Kyung Hee University, Seoul, Republic of Korea}

\begin{abstract}
This study proposes an unsupervised sequence-to-sequence learning approach that automatically assesses the motion-induced reliability degradation of the cardiac volume signal (CVS) in multi-channel electrical impedance-based hemodynamic monitoring. The proposed method attempts to tackle shortcomings in existing learning-based assessment approaches, such as the requirement of manual annotation for motion influence and the lack of explicit mechanisms for realizing motion-induced abnormalities under contextual variations in CVS over time. By utilizing long-short term memory and variational auto-encoder structures, an encoder--decoder model is trained not only to self-reproduce an input sequence of the CVS but also to extrapolate the future in a parallel fashion. By doing so, the model can capture contextual knowledge lying in a temporal CVS sequence while being regularized to explore a general relationship over the entire time-series. A motion-influenced CVS of low-quality is detected, based on the residual between the input sequence and its neural representation with a cut--off value determined from the two-sigma rule of thumb over the training set. Our experimental observations validated two claims: (i) in the learning environment of label-absence, assessment performance is achievable at a competitive level to the supervised setting, and (ii) the contextual information across a time series of CVS is advantageous for effectively realizing motion-induced unrealistic distortions in signal amplitude and morphology. We also investigated the capability as a pseudo-labeling tool to minimize human-craft annotation by preemptively providing strong candidates for motion-induced anomalies. Empirical evidence has shown that machine-guided annotation can reduce inevitable human-errors during manual assessment while minimizing cumbersome and time-consuming processes. The proposed method has a significance especially in the industrial field, where it is unavoidable to gather and utilize a large amount of CVS data to achieve high accuracy and robustness in real-world applications. 
\end{abstract}
\maketitle
	
\section{Introduction}
Multi-channel electrical impedance (MEI)-based cardiopulmonary monitoring has emerged as a promising alternative to conventional technologies (e.g., mechanical ventilation and cardiac catheterization) owing to its invasive nature, which causes discomfort and inconvenience to the subject \cite{borges2012,deibele2008,frerichs2014,kerrouche2001,kubicek1970,zlochiver2006}. MEI measurement is entirely based on several electrodes attached around a human chest and samples temporal data with a fine time-resolution of approximately 0.01-0.02s, such that it is capable of non-invasive, real-time, and long-term continuous monitoring \cite{adler1996,seo2013}. MEI techniques in lung ventilation tracking applications have arrived at a level suitable for commercial and practical use \cite{putensen2019,tomicic2019}; however, it remains in question for hemodynamic monitoring in terms of accuracy, reliability, and long-term persistence \cite{pikkemaat2014}. By virtue of recent endeavors in the bio-impedance fields, accurate and continuous extraction of feeble cardiogenic components in MEI measurement, known as the cardiac volume signal (CVS), is capable of allowing a reliable long-term trace of pivotal hemodynamic quantities, such as stroke volume and cardiac output, in a fully non-invasive manner \cite{jang2020,lee2021}.  

However, motion makes all situations solely different by being a critical troublemaker that considerably interferes with MEI-based measurements \cite{adler1996,dai2008,yang2022,zhang2005}. Owing to the relatively weak cardiogenic impedance change \cite{brown1992,leonhardt2012}, the hemodynamic monitoring system is notably vulnerable, resulting in a significant loss of accuracy and trustworthiness of the extracted CVS \cite{hyun2023}. Although several studies \cite{dai2008,lee2017,soleimani2006,yang2022} have attempted to overcome this hurdle by recovering normal MEI measurements under the motion influence, their effectiveness appears to be uncertain in real-world situations inducing a colossal diversity of motion. Accordingly, as a clinical and industrial compromise within the current MEI technology, there is a preferential demand for sieving motion-influenced CVS of low-quality  \cite{lee2021}. This filtering is intended to warn a device operator of motion corruption in order for minimizing negative ripple effects to the maximum extent, such as a waste of clinical resources and misapprehension regarding health conditions \cite{charlton2021,hyun2023}. 

To this end, an accurate CVS quality-indexing method that assesses motion-induced reliability degradation needs to be developed. Automation and timeliness are crucial as well for the real-time monitoring. To satisfy these requirements, a data-driven solution using machine learning (ML) can be a good fit \cite{belo2017,hyun2021,hyun2020,lecun2015,yun2022}. Hyun \textit{et al.} \cite{hyun2023} recently paved the way for the development of ML-based CVS assessment approaches in a supervised learning framework. Their idea was based on the construction of a barcode-like feature from labeled data, differentiating anomalies from normal CVSs in an individual cardiac cycle. 

\begin{figure}[t]
	\centering
	\includegraphics[width=0.50\textwidth]{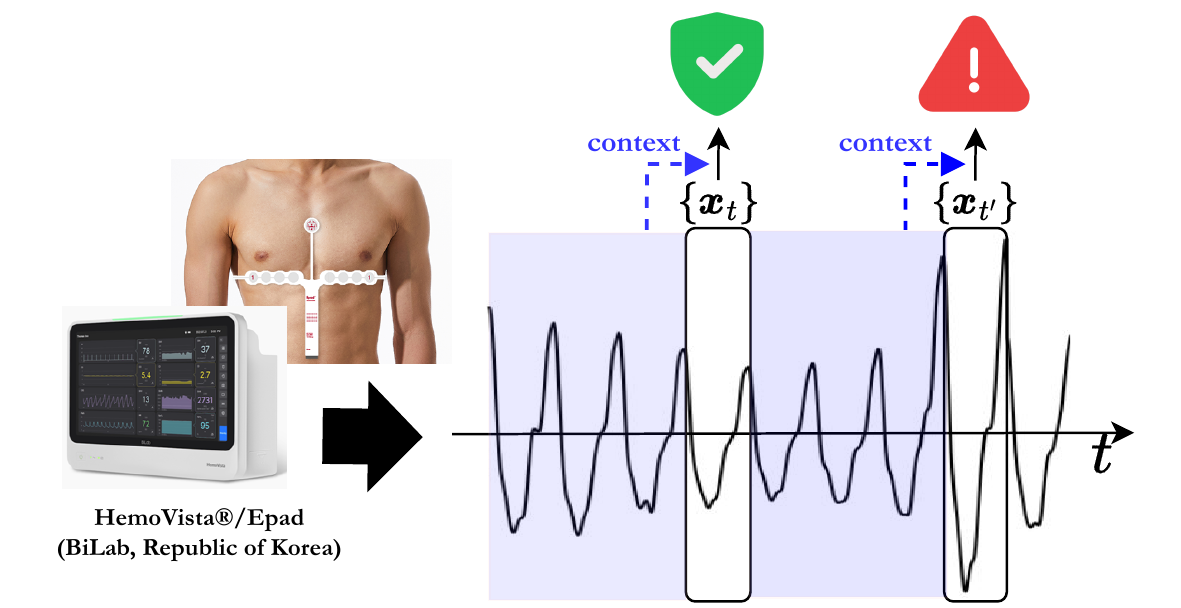}
	\caption{Context knowledge from past is a basis for identifying motion-induced abnormal variations of CVS in real-time monitoring.}
	\label{intro_highlight}
\end{figure}

Despite their remarkable performance, there is yet room for further improvement. The first is the requirement for the manual annotation of motion influence. To enhance model generalization or stability for real-world use, a large amount of data collection and labeling are required; however, they are expensive, time-consuming, and cumbersome \cite{krishnan2022,mao2022}. Moreover, they are prone to inevitable human errors, which may cause biased learning such that the model performance or robustness is limited in practical circumstances \cite{frenay2013,sylolypavan2023,zhu2004}. The second is the lack of explicit mechanisms for realizing motion-induced anomalies under contextual variations in CVS over time. Because of the strong regularity and periodicity according to the heartbeat, as highlighted in Figure \ref{intro_highlight}, the contextual perception across a time-series of CVS is vital, as well as individuals, to identify motion influence, even for bio-signal experts.

To tackle these shortcomings, this study hence proposes an unsupervised sequence-to-sequence learning approach. By utilizing long short-term memory (LSTM) and variational auto-encoder (VAE) structures \cite{hochreiter1997,kingma2013}, an encoder-decoder model is trained not only to self-reproduce an input sequence of CVS but also to extrapolate the future in a parallel fashion. By doing so, the model can capture contextual knowledge lying in a temporal sequence while being regularized to explore a general relationship over a time-series \cite{lin2020,srivastava2015}. The sequence is defined such that its point is either a value of the CVS (point-to-point) or its group during a certain heartbeat interval (cycle-to-cycle), the timing of which is identified from a synchronized electrocardiography (ECG) signal. A motion-influenced CVS of low-quality is detected, based on the residual between an input sequence and its neural representation with a cut--off value determined from the two-sigma rule of thumb \cite{iglewicz1993,pukelsheim1994} over the training set.

The experimental observations validated the following: In a label-absence learning environment, the assessment is achievable at a competitive level in a supervised setting. The best model achieved an accuracy of 0.9566, true positive rate of 0.9743, true negative rate of 0.7001, and AUC of 0.9484, which were comparable to those in the supervised setting, with an accuracy of 0.9672, true positive rate of 0.9847, true negative rate of 0.7151, and AUC of 0.9503. Contextual knowledge across a time series of CVS is advantageous for effectively realizing motion-induced unrealistic distortions in signal amplitude and morphology. In particular, the enriched time context significantly improved the true-negative rate and AUC. Between the two proposed approaches, the cycle-to-cycle model outperforms the point-to-point model. The former achieved an accuracy of 0.9566, true positive rate of 0.9743, true negative rate of of 0.7001, and AUC of 0.9484, whereas the latter an accuracy of 0.9439, true positive rate of 0.9775, true negative rate of 0.4520, and AUC of 0.8338.

We also investigated the capability as a pseudo-labeling tool \cite{lee2013,seibold2022} to minimize human-craft annotation by preemptively providing strong candidates for motion-induced anomalies. Empirical evidence has shown that machine-guided annotation can reduce inevitable human-errors during manual assessment. This shows that the proposed method can synergize with supervised learning as an aide means, not only an alternative branch.

\section{Method}
\begin{figure*}[t]
	\centering
	\includegraphics[width=0.92\textwidth]{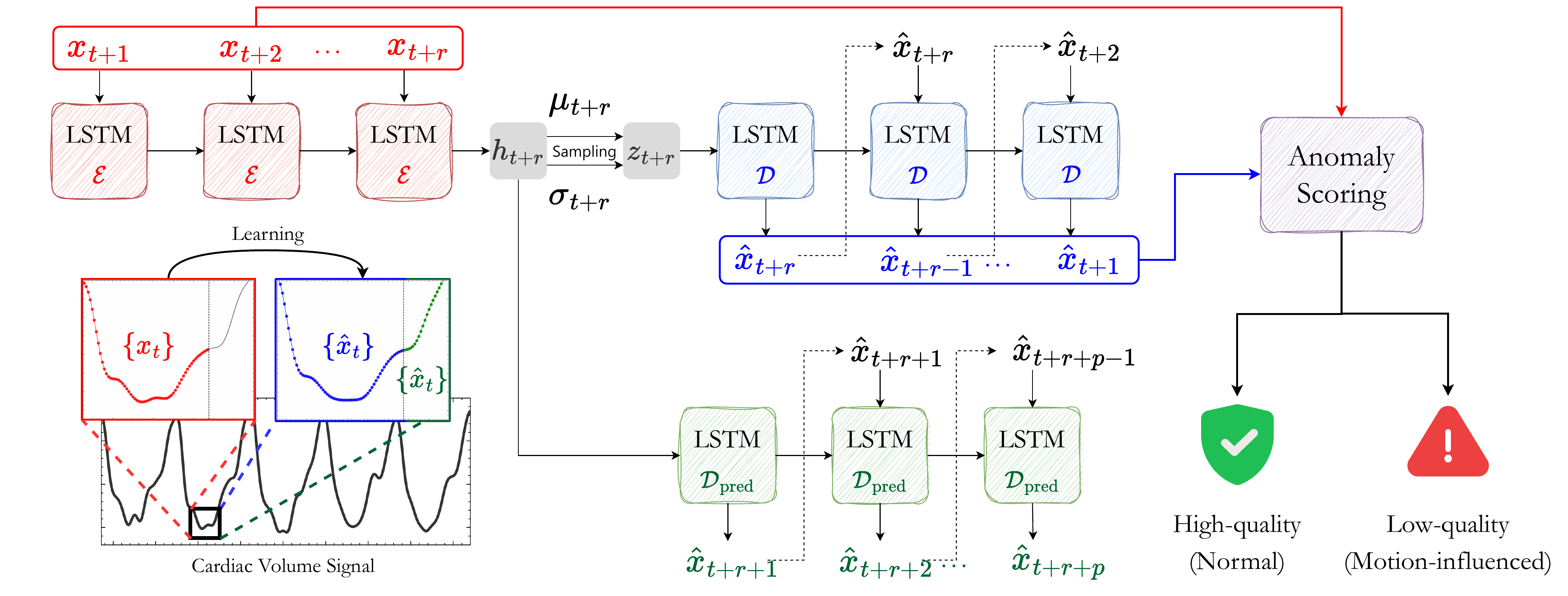}\\
	(a) Point-to-point. \\\vspace{0.1cm}
	\includegraphics[width=0.92\textwidth]{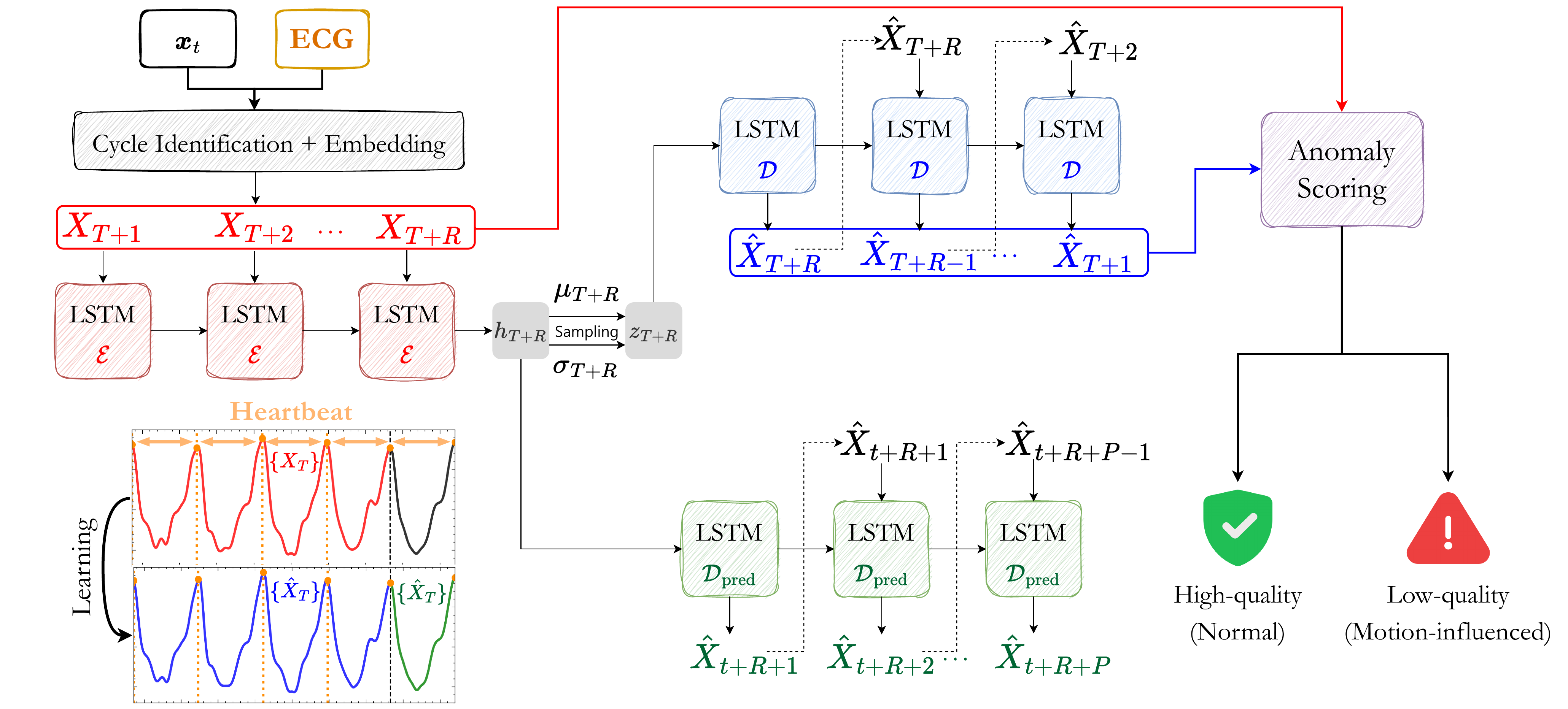}\\
	(b) Cycle-to-cycle.
	\caption{Unsupervised sequence-to-sequence learning models for automatic signal quality assessment in multi-channel impedance-based hemodynamic monitoring: (a) point-to-point and (b) cycle-to-cycle.}
	\label{Proposed}
\end{figure*}
The main objective of this study is to find a CVS quality assessment map $\boldsymbol f:\boldsymbol x_{t} \mapsto \boldsymbol y_{t}$, where $\boldsymbol x_{t}$ is an extracted CVS value from the MEI-based hemodynamic monitoring device \cite{hyun2023,lee2021} at a certain time $t$ and $\boldsymbol y_{t}$ is the corresponding quality index defined by
\begin{equation} \label{idealmap}
	\boldsymbol y_{t} = \boldsymbol f(\boldsymbol x_{t}) = \left\{\begin{array}{cl} 1 & \mbox{if } \boldsymbol x_{t} \mbox{ is normal,} \\ 0 & \mbox{if } \boldsymbol x_{t} \mbox{ is motion-influenced.} \end{array}\right.
\end{equation}
Here, 0 and 1 are numeric values representing low- and high-quality classes, respectively. Considering the previous analysis in \cite{hyun2023}, the transient CVS value $\boldsymbol x_{t}$ can be decomposed as
\begin{equation} \label{motion_relation}
	\boldsymbol x_{t} = \boldsymbol x_{t}^{\mbox{\scriptsize normal}} + \boldsymbol x_{t}^{\mbox{\scriptsize motion}},
\end{equation}
where $\boldsymbol x_{t}^{\mbox{\scriptsize motion}}$ is the motion artifact. See Appendix \ref{appen1} for more details. The map $\boldsymbol f$ in \eqref{idealmap} can be viewed as the identification of $\boldsymbol x_{t}^{\mbox{\scriptsize motion}}$ in $\boldsymbol x_{t}$. To construct $\boldsymbol f$, ML can be directly used in the following supervised learning framework:
\begin{equation}
	\boldsymbol f = \underset{\boldsymbol f}{\mbox{argmin}} ~ \sum_{i} \| \boldsymbol f(\boldsymbol x^{(i)}) - \boldsymbol y^{(i)} \|,
\end{equation}
where $\{(\boldsymbol x^{(i)},\boldsymbol y^{(i)})\}_{i}$ is a paired dataset of the CVS value and corresponding quality index.

However, this approach has two main limitations. The first is a manual annotation. Labeling a large amount of CVS data is extremely costly in terms of human resources and economics. In addition, they are easily exposed to inevitable human errors. One particular source is the ambiguity in the CVS data annotation. For instance, when determining a critical point between normal and motion-influenced time regions, its perfect annotation is almost impossible using only CVS data. Second, point-level identification \eqref{idealmap} is not practically advisable despite relation \eqref{motion_relation}. Even for bio-impedance specialists, the realization of the motion influence is based on contextual knowledge associated with periodicity across repeated cardiac cycles and the time-series recognition of CVS variations. 

\subsection{Unsupervised sequence-to-sequence learning for CVS quality assessment}
The proposed method addresses the aforementioned hurdles. An unsupervised framework is used to learn a barcode-like feature, which plays a key role in sieving the influence of motion, from an unlabeled dataset. The network architecture is a recurrent neural network-style VAE, where LSTM is used as a base building block to provide an explicit mechanism of information propagation that enriches time contextuality. This is motivated by \cite{an2015,lin2020,srivastava2015}.

The proposed method is two-fold: (i) point-to-point and (ii) cycle-to-cycle sequence learning (see Figure \ref{Proposed}). Their main difference lies in the definition of an input sequence sampled from the time-series CVS data. A sequence point is regarded in the former as a CVS value at a certain time and in the latter as a vector gathering all CVS data during a heartbeat interval. Cardiac cycle timing is obtained with the additional use of a synchronized ECG signal. 

\subsubsection{Point-to-point}
\begin{figure}[h]
	\centering
	\includegraphics[width=0.49\textwidth]{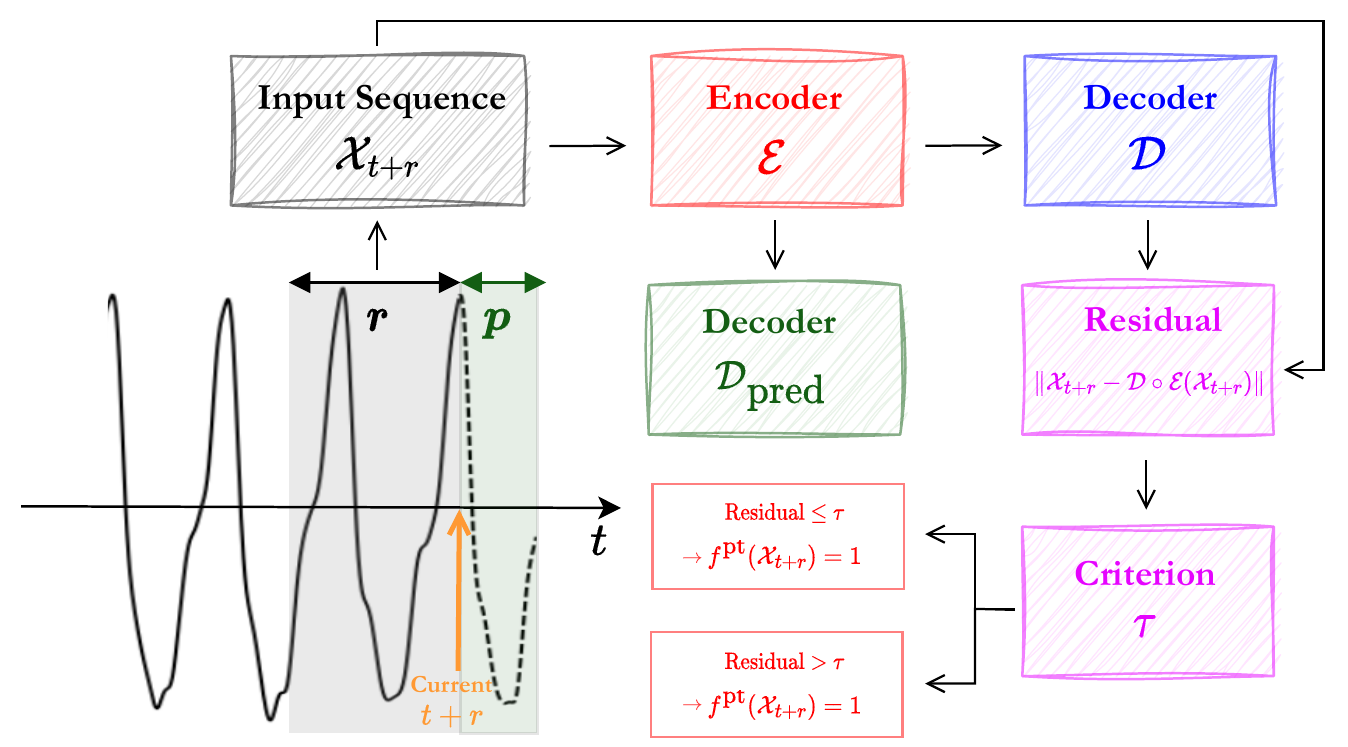}
	\caption{Overall process for the point-to-point model.}
	\label{p2p}
\end{figure}
The point-to-point model $\boldsymbol f^{\mbox{\scriptsize pt}}$ aims to provide  
\begin{equation}\label{pnt_tgt}
	\boldsymbol f^{\mbox{\scriptsize pt}}(\mathcal X_{t+r}) = \boldsymbol y_{t+r},
\end{equation}
where $\mathcal X_{t+r}$ is a consecutive sequence of CVS of length $r$, defined by 
\begin{equation}\label{pointseq_def}
	\mathcal X_{t+r} = \begin{bmatrix} \boldsymbol x_{t+1}, \boldsymbol x_{t+2}, \cdots, \boldsymbol x_{t+r} \end{bmatrix}.
\end{equation}
Here, $t+r$ is set as the current time step for the convenience of notation. The overall process is illustrated in Figure \ref{p2p}. The assessment is performed, based on CVS value histories ($\mathcal X_{t+r}$) to leverage contextual knowledge.

The map $\boldsymbol f^{\mbox{\scriptsize pt}}$ can be expressed as follows:
\begin{equation}
	\boldsymbol f^{\mbox{\scriptsize pt}} = \mathcal T ~ \circ ~ (\mathcal D ~ \circ ~ \mathcal E - \mathcal P),
\end{equation}
where
\begin{itemize}
	\item $\circ$ is the composition operation of functions.
	\item $\mathcal P$ is an operator to reverse the order of vector.
	\item $\mathcal D ~ \circ ~ \mathcal E$ is a VAE-LSTM model.
	\item $\mathcal T$ is an assessment function with a cut--off value $\tau$.
\end{itemize}
Here, the learning process is required for $\mathcal D ~ \circ ~ \mathcal E$. 

Schematically, the encoder $\mathcal E$ and decoder $\mathcal D$ are trained to satisfy
\begin{align}
	\mathcal D ~ \circ ~ \mathcal E (\mathcal X_{t+r})  - \mathcal P (\mathcal X_{t+r}) & \approx \boldsymbol 0, \label{recon}\\
	\mathcal D_{\mbox{\scriptsize pred}} ~ \circ ~ \mathcal E(\mathcal X_{t+r}) & \approx \mathscr X_{t+r,p}, \label{pred}
\end{align}
where $\mathcal D_{\mbox{\scriptsize pred}}$ is another decoder used only for training purposes and $\mathscr X_{t+r,p}$ is the consecutive future sequence of the CVS with a length of $p$, defined by
\begin{equation} \label{prediction}
	\mathscr X_{t+r,p} = \begin{bmatrix} \boldsymbol x_{t+r+1}, \boldsymbol x_{t+r+2}, \cdots, \boldsymbol x_{t+r+p} \end{bmatrix}.
\end{equation}
Here, the decoder $\mathcal D$ reproduces the input sequence $\mathcal X_{t+r}$ in reverse order, as shown in Figure \ref{Proposed} (a). The condition \eqref{recon} causes $\mathcal E$ and $\mathcal D$ to learn a self-representation of the temporal sequence $\mathcal X_{t+r}$. The condition \eqref{pred} applies a regularization force to explore a general relation over the time-series CVS. 

To avoid misleading, we clarify the abuse of notation that $\mathcal E(\mathcal X_{t+r})$ is $\boldsymbol z_{t+r}$ in \eqref{recon} and $\boldsymbol h_{t+r}$ in \eqref{pred}, where $\boldsymbol z_{t+r}$ and $\boldsymbol h_{t+r}$ are defined by
\begin{equation}\label{latent}
	\boldsymbol z_{t+r} = [\mbox{FC}^{\boldsymbol Z}(\boldsymbol Z_{t+r}), \boldsymbol C_{t+r}] \mbox{ and } \boldsymbol h_{t+r}=[\boldsymbol H_{t+r}, \boldsymbol C_{t+r}].
\end{equation}
Here, $\boldsymbol H_{t+r}$ and $\boldsymbol C_{t+r}$ are outputs (hidden and cell states) in the encoder $\mathcal E$ of LSTM, and $\boldsymbol Z_{t+r}$ is given by
\begin{align} 
	&\boldsymbol Z_{t+r} \sim \mathcal N(\boldsymbol \mu_{t+r},\mbox{diag}(\boldsymbol \sigma_{t+r})), \label{sampling} \\
	&\boldsymbol \mu_{t+r} = \mbox{FC}^\mu(\boldsymbol h_{t+r}), \boldsymbol \sigma_{t+r} = \mbox{FC}^\sigma(\boldsymbol h_{t+r}),
\end{align}
where $\mbox{FC}$ is a fully-connected layer with reshaping, $\mbox{diag}(\boldsymbol \sigma)$ is a matrix whose diagonal entries are given by the components of $\boldsymbol \sigma$, and $\mathcal N(\boldsymbol \mu,\Sigma)$ is a Gaussian distribution with a mean of $\boldsymbol \mu$ and covariance of $\boldsymbol \Sigma$. Appendix \ref{appen2} explains more details.

A training objective $J$ is defined as
\begin{align}
	J(\mathcal X_{t+r},\mathscr X_{t+r,p} )  = & ~ \| \mathcal D (\boldsymbol z_{t+r})- \mathcal P (\mathcal X_{t+r}) \|_{\ell_2}^2 \nonumber \\	
	+ &~ \| \mathcal D_{\mbox{\scriptsize pred}}(\boldsymbol h_{t+r}) - \mathscr X_{t+r,p} \|_{\ell_2}^2 \nonumber \\
	+ &~ \mbox{KL}(\mathcal N(\boldsymbol \mu_{t+r},\mbox{diag}(\boldsymbol \sigma_{t+r})) | \mathcal N(\boldsymbol 0 ,\boldsymbol I) ),
\end{align}
where $\mbox{KL}$ is the Kullback--Leibler divergence. The encoder $\mathcal E$ and decoder $\mathcal D$ are optimized in the following sense:
\begin{equation} \label{objective}
	(\mathcal E,\mathcal D,\mathcal D_{\mbox{\scriptsize pred}}) = \underset{(\mathcal E,\mathcal D,\mathcal D_{\mbox{\tiny pred}})}{\mbox{argmin}} ~ \mathbb{E}_{(\mathcal X, \mathscr X)}[\mathcal J(\mathcal X,\mathscr X)],
\end{equation}
where $\mathbb{E}_{(\mathcal X,\mathscr X)}$ is an empirical expectation over ($\mathcal X$,$\mathscr X$). Note that the optimization \eqref{objective} does not involve any loss term associated with the corresponding label $\boldsymbol y$.

The remaining section explains the assessment function $\mathcal T$. We define $\mathcal T$ as
\begin{equation} \label{assessment}
	\mathcal T(\boldsymbol a) = \left\{ \begin{array}{cl} 1 & \mbox{if } ~ \| \boldsymbol a \| \leq \tau \\ 0 & \mbox{if } ~ \| \boldsymbol a \| > \tau  \end{array} \right.,
\end{equation}
where $\| \cdot \|$ is either $\ell_1$ or $\ell_2$ norm. In $\boldsymbol f^{\mbox{\scriptsize pt}}$, $\boldsymbol a$ corresponds to the residual between the original CVS sequence $\mathcal P(\mathcal X)$ and its neural representation $\mathcal D ~ \circ ~ \mathcal E(\mathcal X)$, which is known as an excellent abnormality estimator \cite{an2015,hyun2023}. 

The lingering question is how to determine the cut--off value $\tau$ in \eqref{assessment}. We define a set $\mathbb T_{\tau}$ as
\begin{equation}
	\mathbb T_{\tau} = \left\{ a^{(i)} = \| \mathcal P(\mathcal X^{(i)}) - \widehat{\mathcal X}^{(i)} \| ~ \left| ~ a^{(i)} \leq \tau, i = 1,\cdots,N \right.\right\},
\end{equation}
where $\{ \mathcal X^{(i)} \}_{i=1}^{N}$ is a training dataset of $N$ sequences and $\widehat{\mathcal X}^{(i)} = \mathcal D ~ \circ ~ \mathcal E (\mathcal X^{(i)})$. The cut--off value $\tau$ is determined such that it satisfies the following $\mathbf{2\sigma}$-rule:
\begin{equation}\label{ratio}
	\left \| \begin{array}{l} \underset{\tau \in \mathbb{R}}{\mbox{min}} ~ \tau \\ \mbox{subject to } \dfrac{|\mathbb T_{\tau}|}{N} \geq 0.9545. \end{array} \right.
\end{equation}
Here, $|\cdot|$ is the set cardinality and $0.9545$ can be viewed as the probability that an observation sampled from a normal distribution is within twice standard deviation of the mean.

The point-to-point model assesses the motion influence at each time point, based on the context formed by the learned relationships between CVS values. However, there is still a gap with the experts' perception. Owing to the characteristics of hemodynamic monitoring, a meaningful and salient context is created according to the heartbeat. Although such a context can ideally be learned with large $r$ and $p$, it may be practically restrictive owing to training hurdles associated with very long-term connections and high learning complexity. The cycle-to-cycle model is then modeled to more similarly mimic the heuristic perception of bio-signal experts, who strongly take advantage of contexts associated with heartbeat-related regularity and periodicity. 

\subsubsection{Cycle-to-cycle}
\begin{figure}[h]
	\centering
	\includegraphics[width=0.49\textwidth]{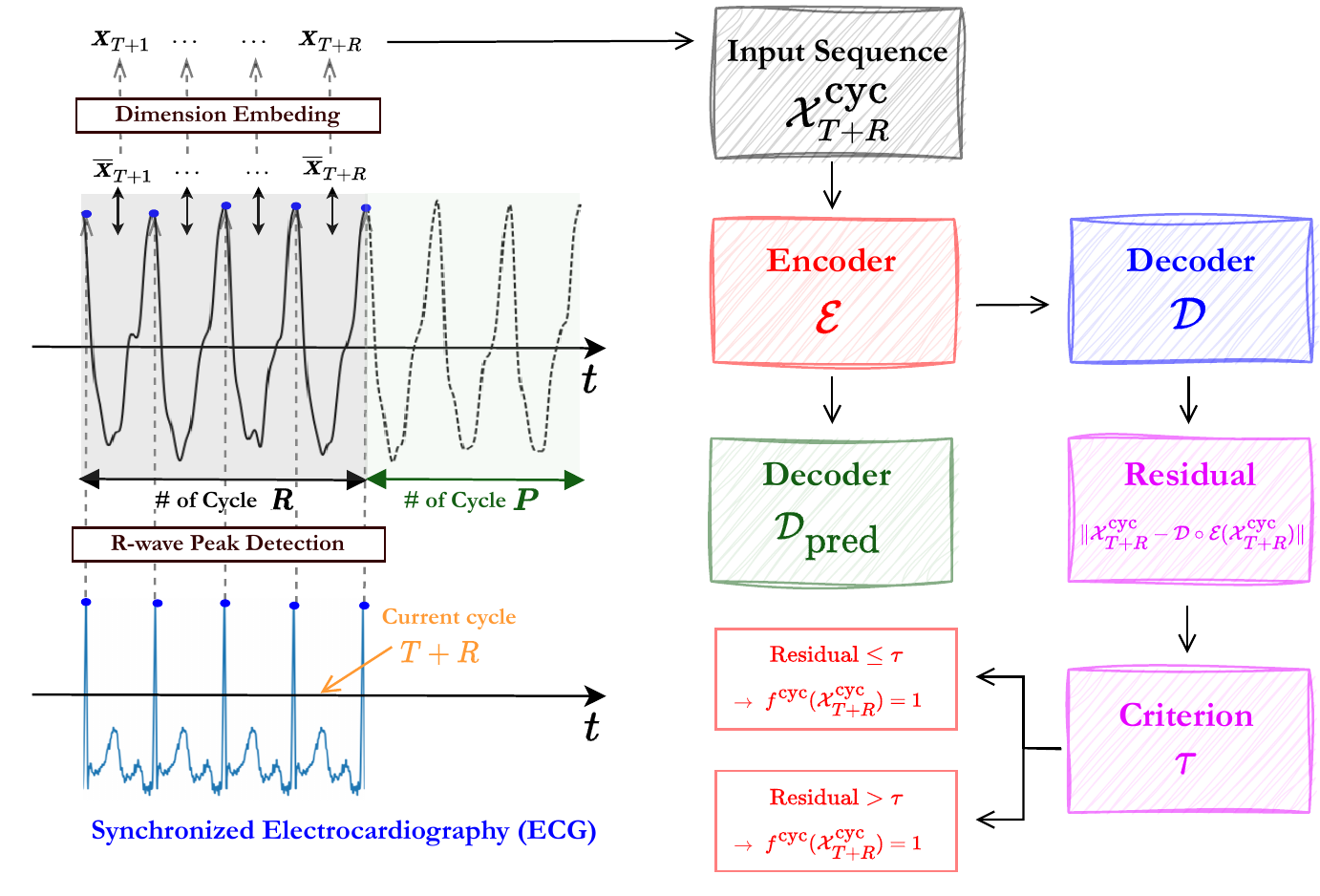}
	\caption{Overall process for the cycle-to-cycle model.}
	\label{c2c}
\end{figure}
The cycle-to-cycle model $\boldsymbol f^{\mbox{\scriptsize cyc}}$ aims to provide
\begin{equation} \label{cyctarget}
	\boldsymbol f^{\mbox{\scriptsize cyc}}(\mathcal X_{T+R}^{\mbox{\scriptsize cyc}}) = \boldsymbol Y_{T+R},
\end{equation}
where $\mathcal X_{T+R}^{\mbox{\scriptsize cyc}}$ is a consecutive CVS sequence with a length of $R$, defined by 
\begin{equation}\label{defcycX}
	\mathcal X_{T+R}^{\mbox{\scriptsize cyc}} = \begin{bmatrix} \overline{\boldsymbol X}_{T+1}, \overline{\boldsymbol X}_{T+2}, \cdots, \overline{\boldsymbol X}_{T+R}\end{bmatrix},
\end{equation}
and $\boldsymbol Y_{T+R} \in \{0,1\}$ is the corresponding assessment labels. Here, $\overline{\boldsymbol X}_{T}$ represents a vector gathering all the CVS values during the $T$-th cardiac cycle. The assessment is based on accumulated cardiac cycle histories. 

However, there here are two uncertainties: (i) identification of a cardiac cycle from consecutive times and (ii) addressing inconsistent point dimensions caused by the nature of heart-rate variability \cite{hyun2023,shaffer2017}. Definition \eqref{defcycX} is informal because the dimension of $\overline{\boldsymbol X}_{T}$ does not usually match that of $\overline{\boldsymbol X}_{T^\prime}$ for $T^\prime \neq T$.
 
The structure of $\boldsymbol f^{\mbox{\scriptsize cyc}}$ is conceptually equivalent to that of point-to-point, whereas an additional preprocessing $\mathcal C$ is employed to resolve the aforementioned issues. $\boldsymbol f^{\mbox{\scriptsize cyc}}$ can be expressed as follows:
\begin{equation}\label{cyc}
	\boldsymbol f^{\mbox{\scriptsize cyc}} = \mathcal T ~ \circ ~ (\mathcal D ~ \circ ~ \mathcal E - \mathcal P) ~ \circ ~ \mathcal C.
\end{equation}
Here, the pre-processing $\mathcal C$ provides an input CVS sequence $\mathcal X_{T+R}^{\mbox{\scriptsize cyc}}$ from ECG and CVS data (see Figure \ref{c2c}).

The detailed procedures of $\mathcal C$ are as follows: From the synchronized ECG signal data, we first identify the timing of the $(T+R)$-th cardiac cycle through R-wave peak detection \cite{manikandan2012}. Thereafter, we obtain $\overline{\boldsymbol X}_{T+R}$ and interpolate it such that having a fixed dimension. Denoting the interpolated vector as $\boldsymbol X_{T+R}$, we obtain 
\begin{equation}
	\mathcal X_{T+R}^{\mbox{\scriptsize cyc}} = \begin{bmatrix} \boldsymbol X_{T+1}, \boldsymbol X_{T+2}, \cdots, \boldsymbol X_{T+R}\end{bmatrix},
\end{equation}
In our implementation, linear interpolation was used and the embedding dimension was $150$.

The encoder--decoder model $\mathcal D ~ \circ ~ \mathcal E$ is trained with the aid of $\mathcal D_{\mbox{\scriptsize pred}}$ in the same manner as in \eqref{objective}, where $\mathcal D_{\mbox{\scriptsize pred}}$ is trained to estimate a future sequence $\mathscr X_{T+R,P}^{\mbox{\scriptsize cyc}}$ with a length of $P$, defined by
\begin{equation}
	\mathscr X_{T+R,P}^{\mbox{\scriptsize cyc}} = \begin{bmatrix} {\boldsymbol X}_{T+R+1}, \boldsymbol X_{T+R+2}, \cdots, \boldsymbol X_{T+R+P} \end{bmatrix}.
\end{equation}
The final assessment $\mathcal T$ is in the same manner as \eqref{assessment}.

\section{Experiments and Results}
\subsection{Experimental Set-up}
To evaluate the performance of the proposed method, we used a labeled dataset sourced from \cite{hyun2023}, in which CVS and synchronized ECG data were obtained from 19 healthy subjects using a commercial MEI-based hemodynamic monitoring device (HemoVista, BiLab, Republic of Korea). For the cardiac cycle, the dataset comprises total 12928 normal and 3212 motion influenced cycles. We have clearly mentioned that annotated labels were only used for model performance comparison.

We conducted ML experiments in a computing environment with four GeForce RTX 3080 Ti devices, two Intel Xeon CPUs E5-2630 v4, and 128GB DDR4 RAM. Our implementation was based on PyTorch \cite{paszke2019} and PyTorch-lightning. See Appendix \ref{appen2} for network and training details.

For quantitative analysis, the following evaluation metrics (ACC, TPR, TNR, and AUC) were used, where
\begin{align}
	\mbox{ACC}=\dfrac{N_{\mbox{\scriptsize TP}}+N_{\mbox{\scriptsize TN}}}{N_{\mbox{\scriptsize TP}}+N_{\mbox{\scriptsize TN}}+N_{\mbox{\scriptsize FP}}+N_{\mbox{\scriptsize FN}}}, \\ \mbox{TPR}=\dfrac{N_{\mbox{\scriptsize TP}}}{N_{\mbox{\scriptsize TP}}+N_{\mbox{\scriptsize FP}}}, \mbox{TNR}=\dfrac{N_{\mbox{\scriptsize TN}}}{N_{\mbox{\scriptsize TN}}+N_{\mbox{\scriptsize FN}}},
\end{align}
and AUC is area under the curve of receiver operating characteristic (ROC). Here, $N_{\mbox{\scriptsize TP}}$, $N_{\mbox{\scriptsize TN}}$, $N_{\mbox{\scriptsize FP}}$, and $N_{\mbox{\scriptsize FN}}$ are the numbers of true positives, true negatives, false positives, and false negatives, respectively. Because our dataset was highly imbalanced, TNR and AUC were emphasized much more than ACC and TPR.

\subsection{Results}
\subsubsection{Point-to-point}
This subsection demonstrates experimental results for the point-to-point approach. 

\paragraph{Model performance} 
\begin{table}[t]
	\centering
	\begin{tabular}{p{0.5cm}p{0.5cm}|p{1cm}p{1cm}p{1cm}p{1cm}}
		\textbf{r} & \textbf{p} & \textbf{ACC} & \textbf{TPR} & \textbf{TNR} & \textbf{AUC}  \\
		\hline\hline 1 & 0 & 0.9105 & 0.9595 & 0.1894 & 0.6009 \\
		1 & 1 & 0.9125 & 0.9605 & 0.2054 & 0.6064 \\
		\hline 50 & 0 & 0.9264 & 0.9681 & 0.3152 & 0.7044 \\
		50 & 50 & 0.9359 & 0.9730 & 0.3889 & 0.7747 \\
		\hline 100 & 0 & 0.9308 & 0.9703 & 0.3490 & 0.7717 \\
		100 & 100 & 0.9391 & 0.9748 & 0.4143 & 0.8084 \\
		\hline 150 & 0 & 0.9316 & 0.9708 & 0.3554 & 0.7764 \\
		150 & 150 & 0.9405 & 0.9756 & 0.4259 & 0.8227 \\
		\hline 200 & 0 & 0.9361 & 0.9733 & 0.3914 & 0.7910 \\
		200 & 50 & 0.9376 & 0.9740 & 0.4030 & 0.7954 \\
		200 & 150 & 0.9357 & 0.9747 & 0.4116 & 0.8052 \\
		\textbf{200} & \textbf{200} & \textbf{0.9439} & \textbf{0.9775} & \textbf{0.4520} & \textbf{0.8338} \\
		200 & 300 & 0.9351 & 0.9728 & 0.3839 & 0.7858 \\
        \hline 300 & 300 & 0.9367 & 0.9737 & 0.3966 & 0.8167 \\
	\end{tabular}
	\caption{Quantitative performance for the point-to-point models.}
	\label{perfom_quanti}
\end{table}

Table \ref{perfom_quanti} summarizes the quantitative evaluation results for varying the reconstruction sequence length $r$ and prediction $p$. The empirically best performance was achieved for $r=p=200$, which included points of approximately 1.5 heartbeat cycles in the reconstruction and prediction sequences. For a given $r$, $p \approx r$ tends to produce a higher performance than $p=0$ in terms of all statistical metrics. In the case of $r=200$, as $p$ increased, the model provided improved outcomes owing to enriched contextual information, whereas the model performance degraded for $p=300$ because of the increased learning complexity or over-regularization. In our empirical implementation, the prediction decoder played a key role in stabilizing the model training and improving the ability to assess the motion influence.

\begin{figure}[t]
	\centering
	\includegraphics[width=0.49\textwidth]{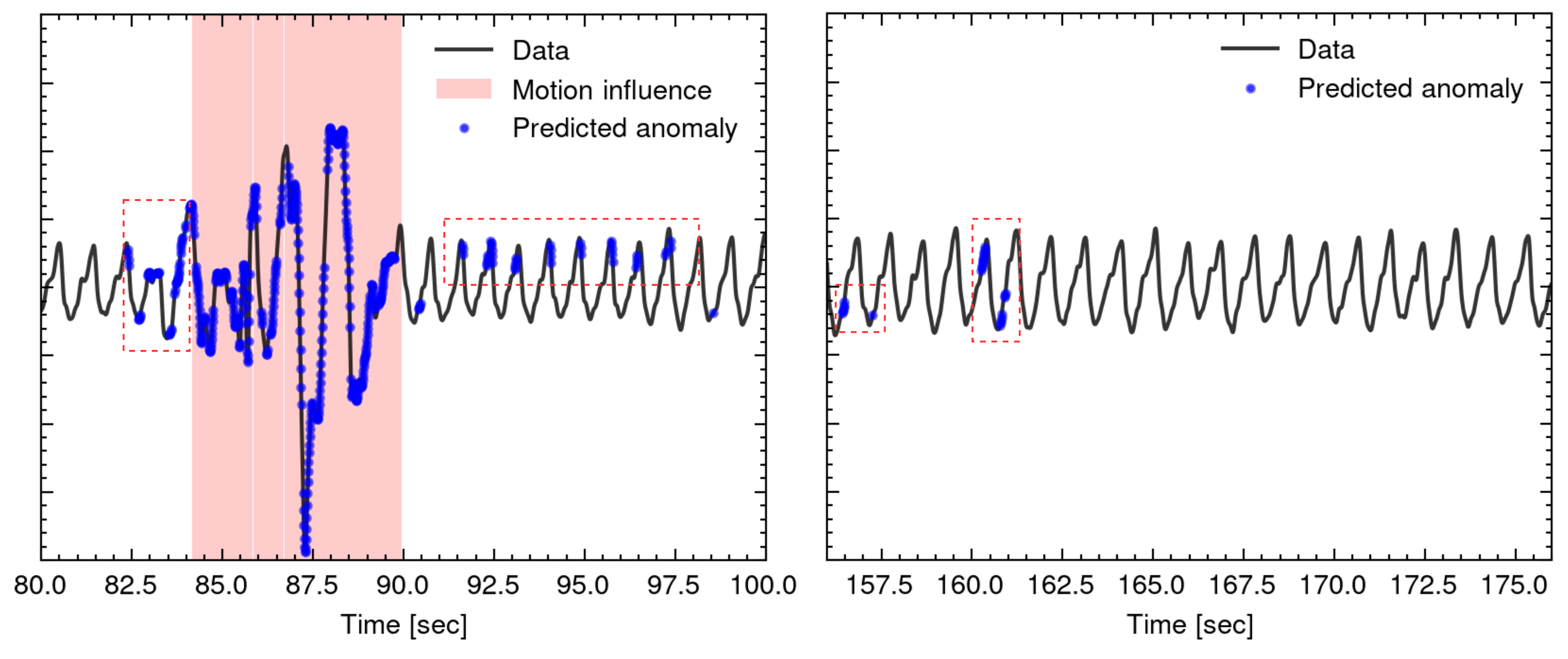}
	\caption{Qualitative performance for the point-to-point model with $r=p=200$.}
	\label{perfom_quali}
\end{figure}
\begin{figure}[t]
	\centering
	\includegraphics[width=0.375\textwidth]{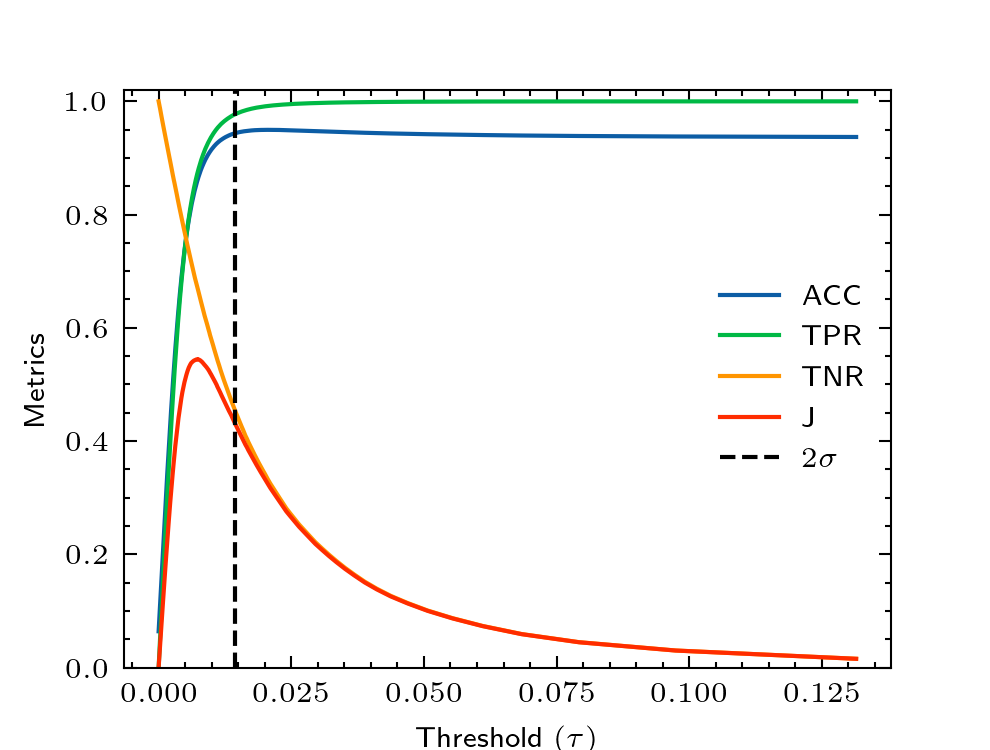} \vspace{0.3cm}\\
	\begin{tabular}{p{1cm}|p{1cm}p{1cm}p{1cm}p{1cm}}
		\textbf{cut--off} & \textbf{ACC} & \textbf{TPR} & \textbf{TNR} & \textbf{AUC} \\ \hline
		$\mathbf{2\sigma}$ & 0.9439 & 0.9775 & 0.4520 & \multirow{2}{*}{0.8338} \\
		$\boldsymbol J_{\mbox{\scriptsize max}}$ & 0.8591 & 0.8718 & 0.6726 & \\ 
	\end{tabular}
	\caption{Quantitative evaluation by changing $\tau$ for the point-to-point model with $r=p=200$.}
	\label{p2p_cutoff}
\end{figure}

Figure \ref{perfom_quali} shows the qualitative evaluation. This point-to-point approach has several limitations. In the left case, the model appeared to predict abnormal regions well; however, point-level mis-identifications were frequently observed, which did not occur from a heuristic perspective (see the red dotted boxes). The case shown on the right presents a similar problem. These results motivated us to introduce the cycle-to-cycle approach.

\begin{table}[t]
	\centering
	\begin{tabular}{p{0.5cm}p{0.5cm}|p{1cm}p{1cm}p{1cm}p{1cm}}
		\textbf{R} & \textbf{P} & \textbf{ACC} & \textbf{TPR} & \textbf{TNR} & \textbf{AUC} \\
		\hline\hline 1 & 0 & 0.9532 & 0.9748 & 0.6419 & 0.9158 \\
		1 & 1 & 0.9590 & 0.9813 & 0.6294 & 0.9251 \\
		\hline 2 & 0 & 0.9580 & 0.9805 & 0.6309 & 0.9232 \\			
		2 & 1 & 0.9545 & 0.9748 & 0.6600 & 0.9450 \\
		\textbf{2} & \textbf{2} & \textbf{0.9566} & \textbf{0.9743} & \textbf{0.7001} & \textbf{0.9484} \\
		2 & 3 & 0.9567 & 0.9754 & 0.6861 & 0.9353 \\
		\hline 3 & 3 & 0.9659 & 0.9868 & 0.6220 & 0.9381 \\
	\end{tabular}
	\caption{Quantitative performance for the cycle-to-cycle models.}
	\label{c2c_perfom_quanti}
\end{table}
\begin{figure}[t]
	\centering
	\includegraphics[width=0.49\textwidth]{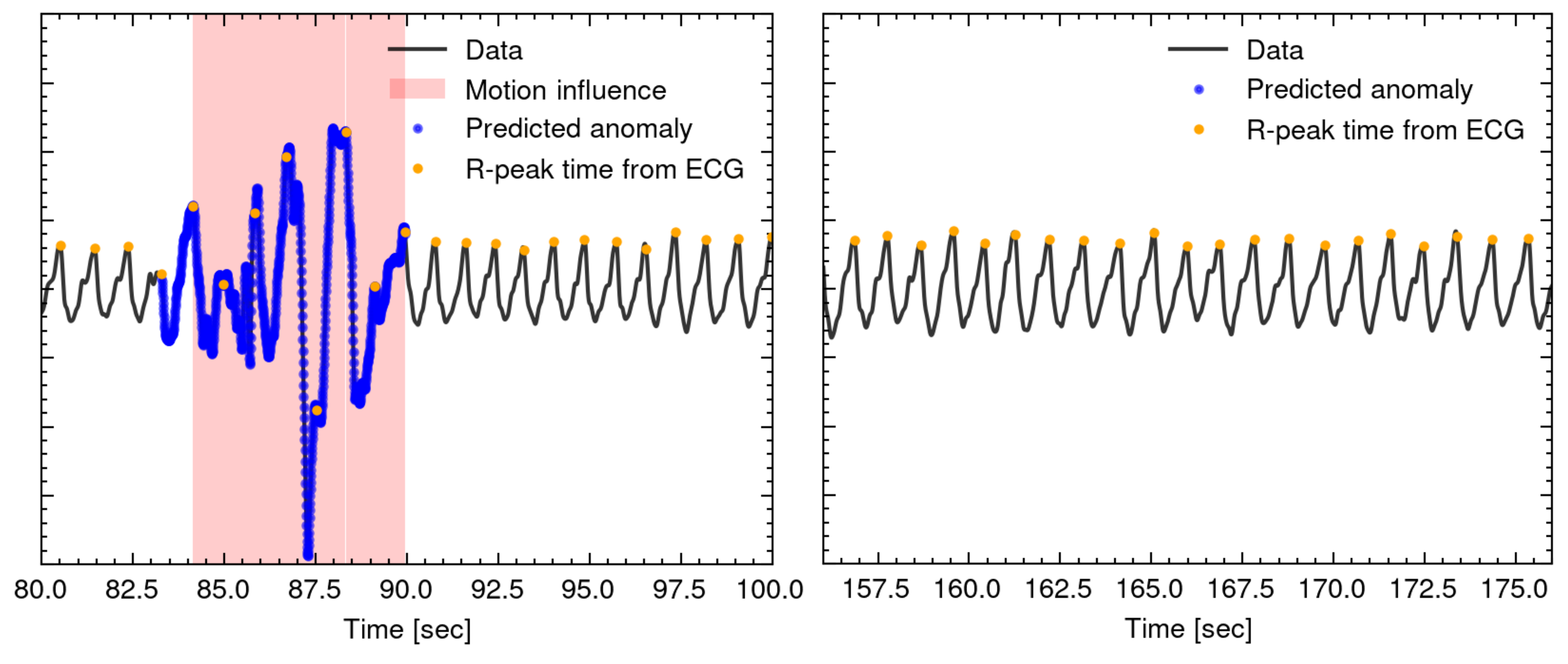}
	\caption{Qualitative performance for the cycle-to-cycle model with $R=P=2$.}
	\label{c2c_perfom_quali}
\end{figure}
\begin{figure}[t]
	\centering
	\includegraphics[width=0.375\textwidth]{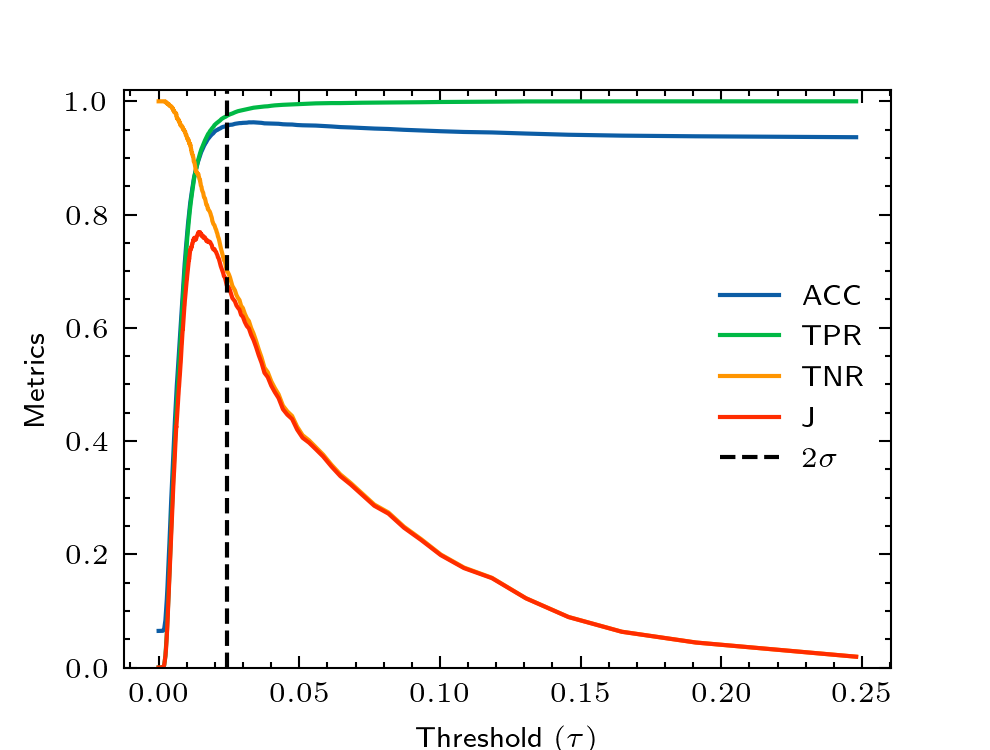} \vspace{0.3cm}\\
	\begin{tabular}{p{1cm}|p{1cm}p{1cm}p{1cm}p{1cm}}
		\textbf{cut--off} & \textbf{ACC} & \textbf{TPR} & \textbf{TNR} & \textbf{AUC}  \\ \hline
		$\mathbf{2\sigma}$ & 0.9566 & 0.9743 & 0.7001 & \multirow{2}{*}{0.9484}  \\
		$\boldsymbol J_{\mbox{\scriptsize max}}$ & 0.8966 & 0.8984 & 0.8706 & \\	\end{tabular}
	\caption{Quantitative evaluation by changing $\tau$ for the cycle-to-cycle model with $R=P=2$.}
	\label{c2c_cutoff}
\end{figure}
\begin{figure*}[p]
	\centering
	\includegraphics[width=0.99\textwidth]{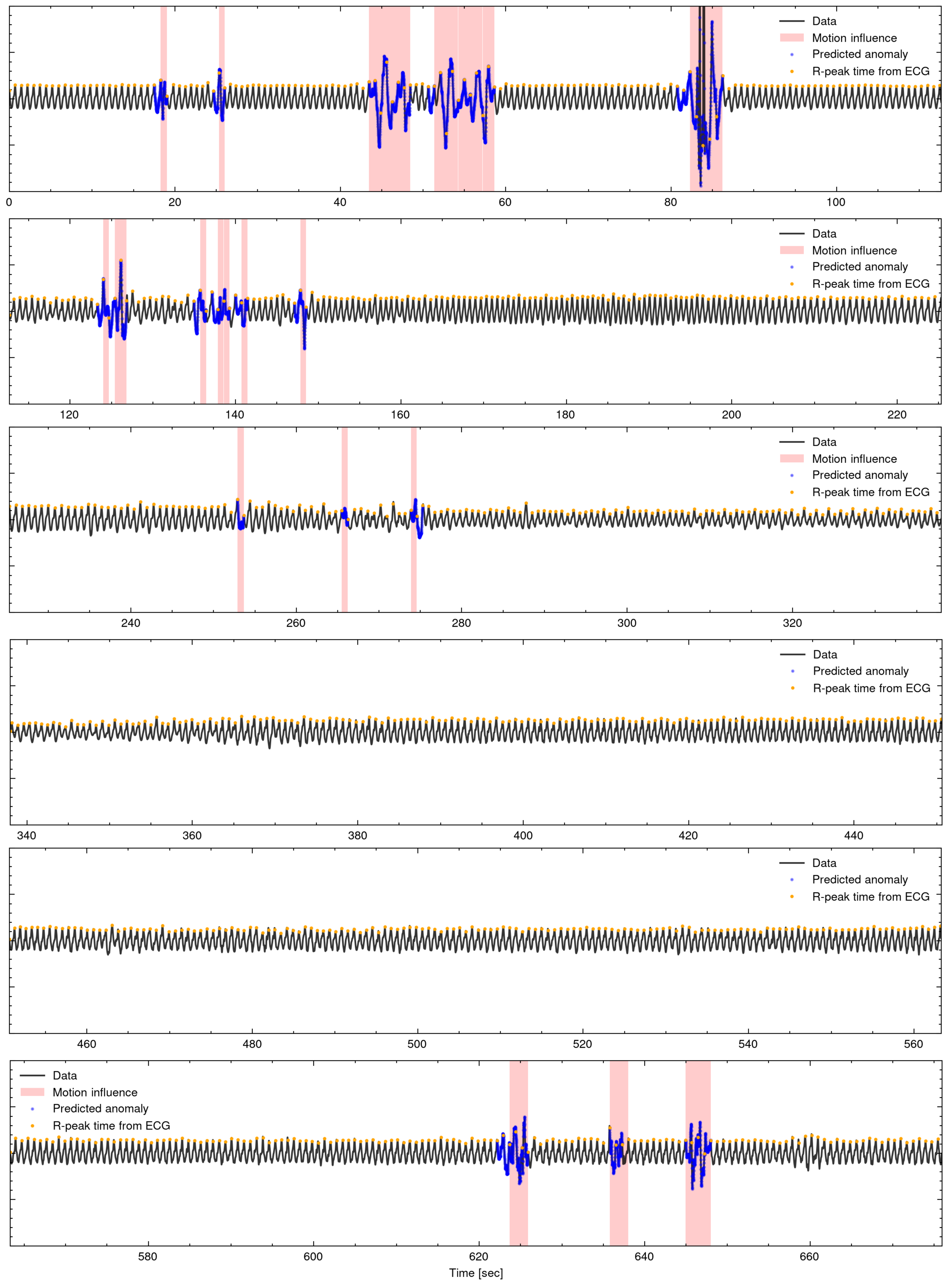}
	\caption{Qualitative assessment results using the cycle-to-cycle method with $R=P=2$ for all the time-series data from one test subject.}
	\label{entire_eval}
\end{figure*}

\paragraph{About cut--off value $\tau$} 
We quantitatively compared the model performance by varying the cut--off value $\tau$ (see Figure \ref{p2p_cutoff}). We introduced Youden's $J$-statistic as a new evaluation metric, which is defined as
\begin{equation}
	J = \mbox{TPR} + \mbox{TNR} - 1
\end{equation}
Because $J$ is known to be an excellent indicator for determining an optimal cut--off value in a class-imbalanced dataset \cite{ruopp2008}, the proposed method was compared to the model at the cut--off value $\tau$ that maximizes $J$. Here, we clarify that $J$ cannot be obtained in an unsupervised setting. The empirical results in Figure \ref{p2p_cutoff} support the claim that the two-sigma rule of thumb in \eqref{ratio} produces a competitive choice. 

\begin{figure}[t]
	\centering
	\includegraphics[width=0.47\textwidth]{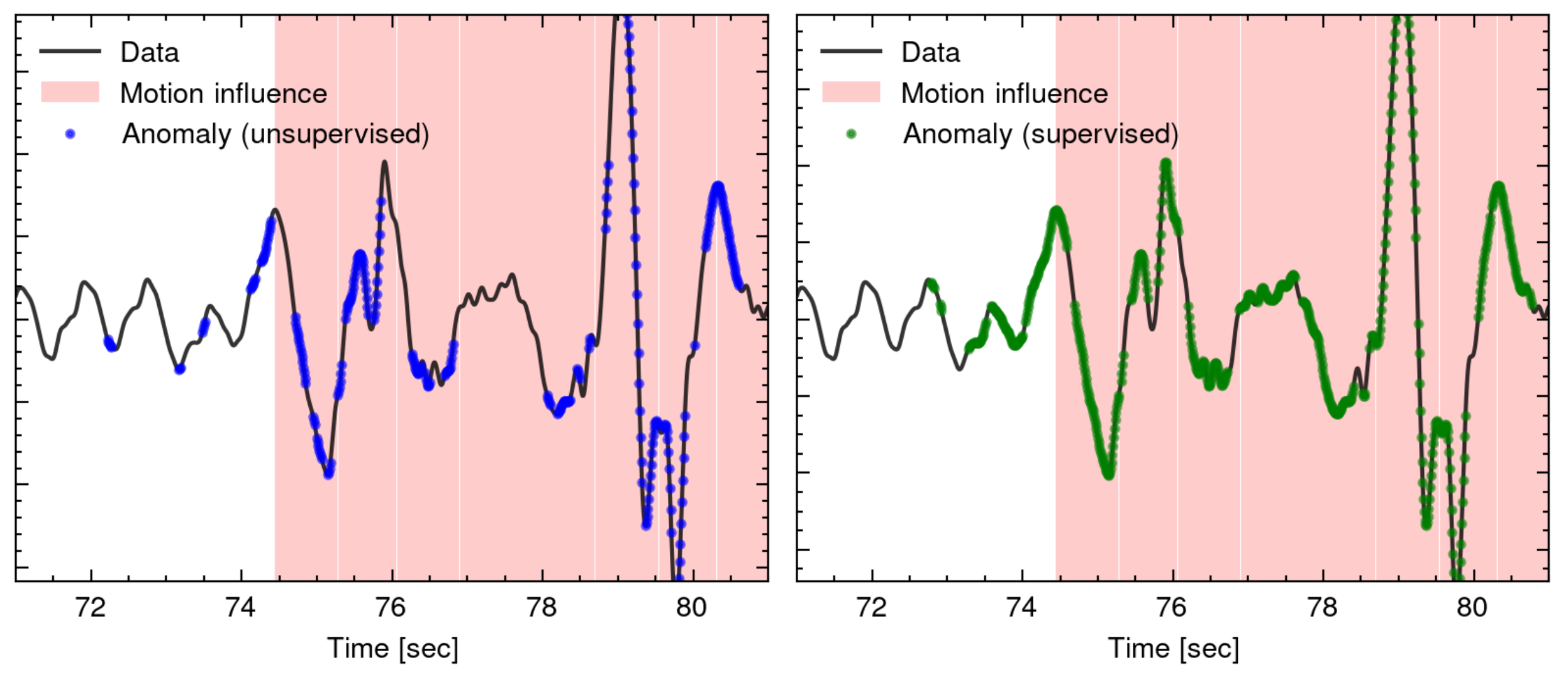} \\ \vspace{0.2cm}
	\begin{tabular}{p{0.5cm}|p{1cm}|p{1.0cm}p{1.0cm}p{1.0cm}p{1.0cm}} 
		& \textbf{cut--off} & \textbf{ACC} & \textbf{TPR} & \textbf{TNR} & \textbf{AUC} \\ \hline
		\multirow{2}{*}{\textbf{SL}} & $\mathbf{2\sigma}$  & 0.9463 & 0.9787 & 0.4707 & \multirow{2}{*}{0.8690} \\ 
		& $\boldsymbol J_{\mbox{\scriptsize max}}$ & 0.8677 & 0.8764 & 0.7397 & \\ \hline 
		\multirow{2}{*}{\textbf{USL}} & $\mathbf{2\sigma}$ & 0.9439 & 0.9775 & 0.4520 & \multirow{2}{*}{0.8338} \\ 
		&  $\boldsymbol J_{\mbox{\scriptsize max}}$ & 0.8591 & 0.8718 & 0.6726 &  \\
	\end{tabular} \\ \vspace{0.1cm}
	(a) Point-to-point \vspace{0.1cm} \\
	\includegraphics[width=0.47\textwidth]{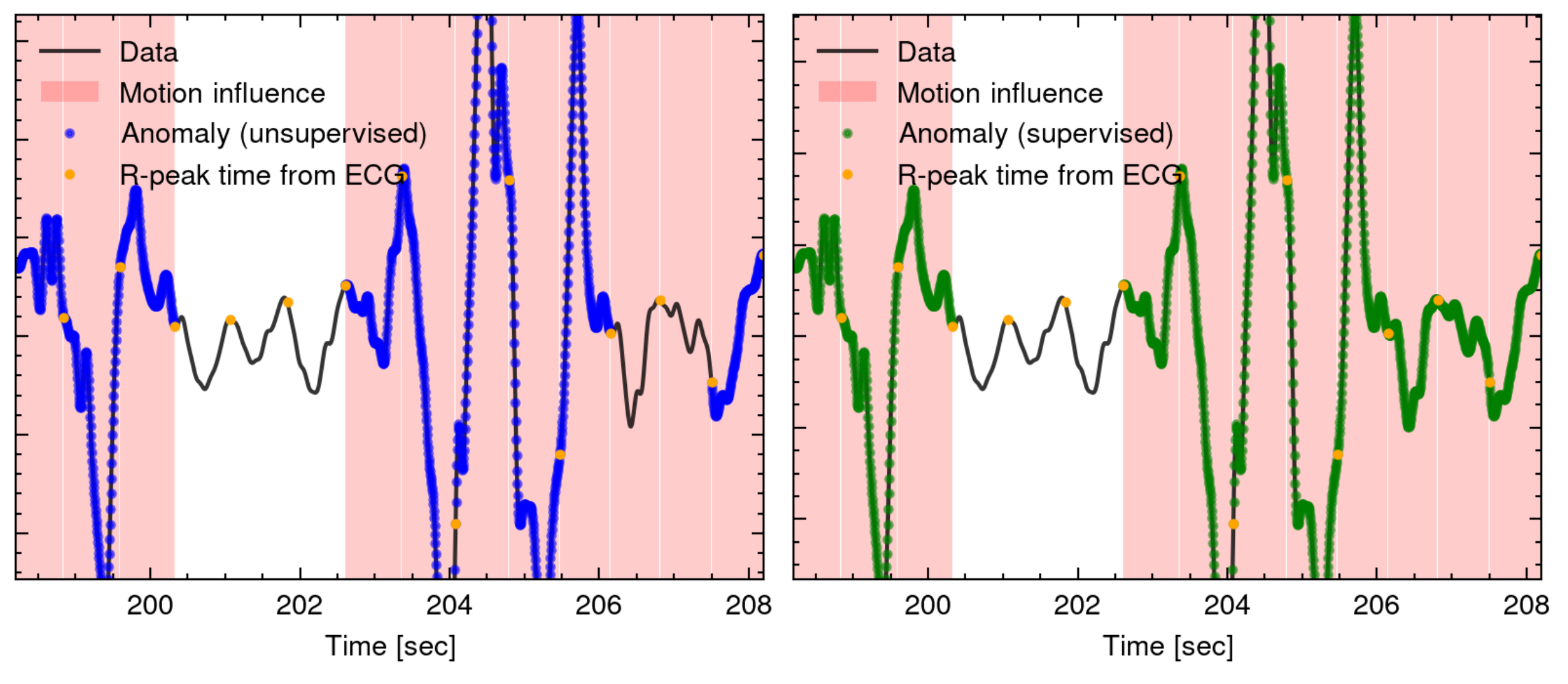} \vspace{0.1cm} \\
	\begin{tabular}{p{0.5cm}|p{1cm}|p{1.0cm}p{1.0cm}p{1.0cm}p{1.0cm}} 
		& \textbf{cut--off} & \textbf{ACC} & \textbf{TPR} & \textbf{TNR} & \textbf{AUC} \\ \hline
		\multirow{2}{*}{\textbf{SL}} & $\mathbf{2\sigma}$ & 0.9658 & 0.9829 & 0.7182 & \multirow{2}{*}{0.9630} \\ 
		& $\boldsymbol J_{\mbox{\scriptsize max}}$ & 0.9260 & 0.9288 & 0.8847 & \\ \hline 
		\multirow{2}{*}{\textbf{USL}} & $\mathbf{2\sigma}$ & 0.9566 & 0.9743 & 0.7001 & \multirow{2}{*}{0.9484} \\ 
		& $\boldsymbol J_{\mbox{\scriptsize max}}$ & 0.8966 & 0.8984 & 0.8706 & \\ 
	\end{tabular} \\ \vspace{0.1cm}
	(b) Cycle-to-cycle
	\caption{Quantitative and qualitative comparison in supervised and unsupervised settings: (a) point-to-point ($r=p=200$) and (b) cycle-to-cycle ($R=P=2$) models.}
	\label{splvsuspl}
\end{figure}

\subsubsection{Cycle-to-cycle}
This subsection demonstrates experimental results for the cycle-to-cycle approach. 

\paragraph{Model performance} 
Table \ref{c2c_perfom_quanti} summarizes the quantitative evaluation results for varying $R$ and $P$. The overall performance was significantly better than that of the point-to-point. The best empirical model is of $R=P=2$. Similar to the point-to-point, the use of prediction was empirically advantageous to enhance the training stability and final CVS assessment performance. The qualitative evaluation results are exhibited in Figure \ref{c2c_perfom_quali}. The cycle-to-cycle model successfully addressed the limitation in the point-to-point. 

\paragraph{About cut--off value $\tau$} 
In Figure \ref{c2c_cutoff}, model performance was quantitatively compared by varying $\tau$. In the cycle-to-cycle model, the two-sigma rule of thumb provided an outcome very close to the optimal in terms of $J$ statistic. 

Figure \ref{entire_eval} visualizes CVS qualitative assessment results of the cycle-to-cycle model for all the time-series data from one test subject.

\subsubsection{Comparison with supervised learning setting}
This subsection examines the extent to which the performance gap exists between supervised and unsupervised settings. By utilizing only positive samples from the labeled dataset, the proposed models were trained to learn an underlying low dimensional distribution from motion-free CVS sequence data, which is desired to be implicitly realized in the unsupervised set-up. 

Figure \ref{splvsuspl} exhibits the corresponding quantitative and qualitative evaluation results for the (a) point-to-point and (b) cycle-to-cycle models. Even though supervised learning was superior, unsupervised learning provided comparable outcomes in a label-absence environment.

\subsection{Investigation as a pseudo-labeling tool} \label{PLT}
\begin{figure}[t]
	\centering
	\includegraphics[width=0.5\textwidth]{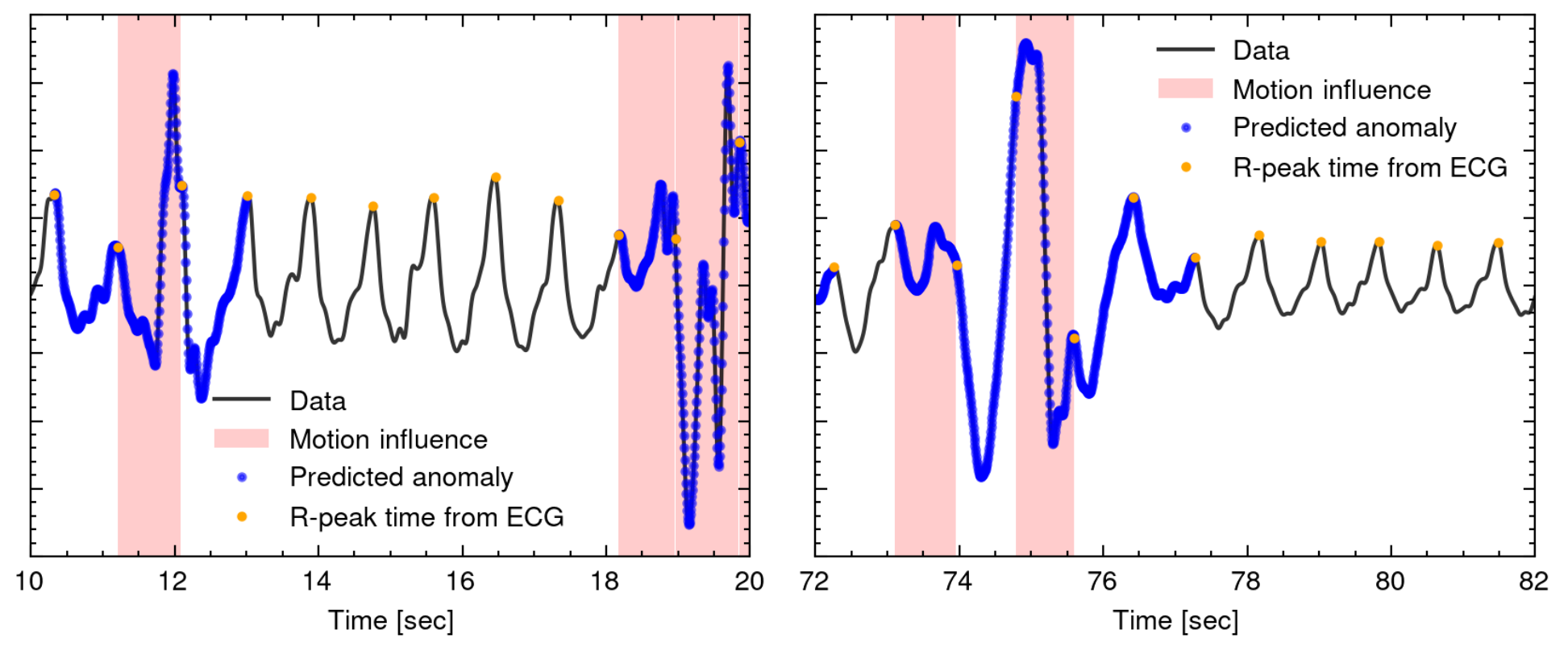} \\
	(a) Original Annotation \vspace{0.2cm} \\ 
	\includegraphics[width=0.5\textwidth]{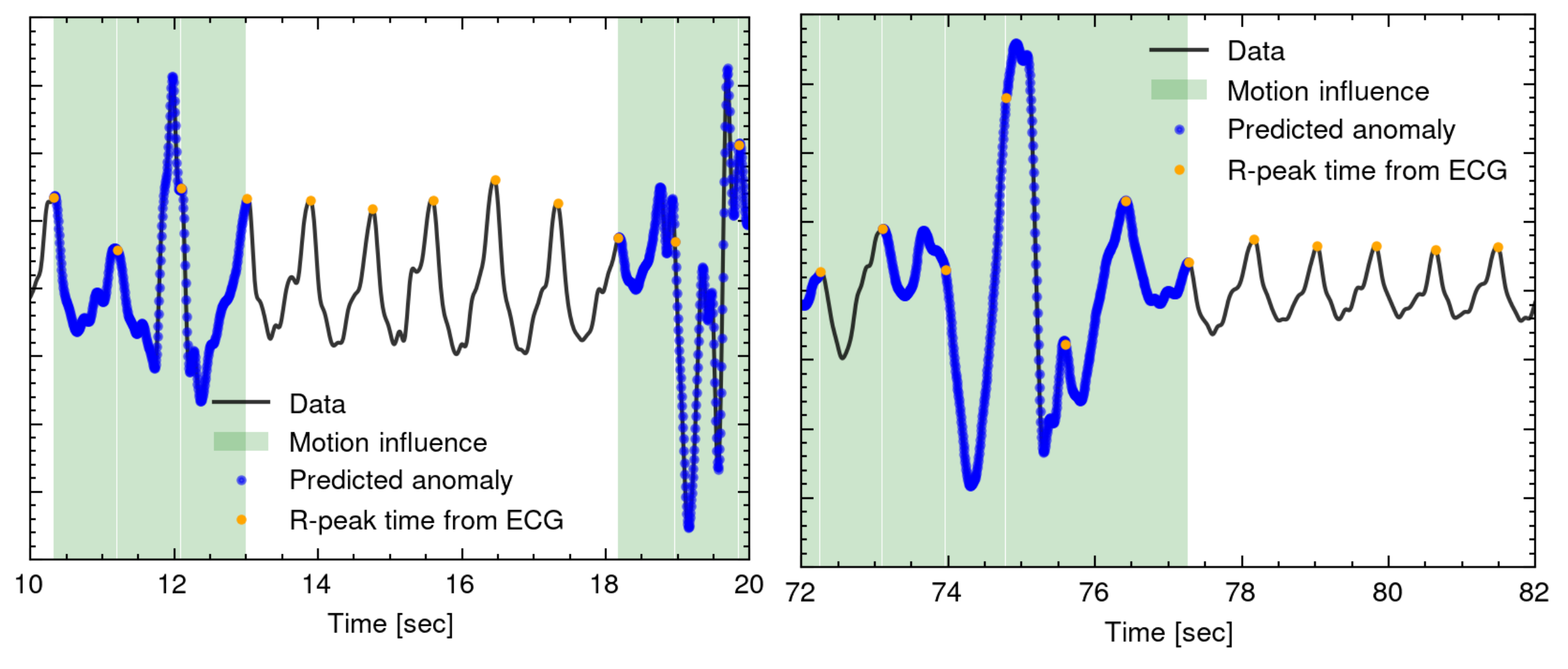} \\
	(b) Machine-guided Annotation \vspace{0.2cm} \\
	\begin{tabular}{p{15mm}|p{15mm}p{15mm}p{15mm}} 
		& \textbf{original} & \textbf{prediction} & \textbf{guided} \\ \hline
		\textbf{original} & 0 & 0.0800 & 0.0253 \\ 
		\textbf{prediction} & 0.0800 & 0 & 0.0695 \\ 
		\textbf{guided} & 0.0253 & 0.0695 & 0 \\
	\end{tabular} \vspace{0.2cm} \\ 
	(c) Inter-observability \vspace{0.2cm} \\
	\caption{Qualitative and quantitative comparisons among original human-craft annotation (red regions), machine-driven prediction (blue dots), and machine-guided annotation (green regions).}
	\label{pltool}
\end{figure}

This section investigates our method as a pseudo-labeling tool that can not only reduce time-consuming processes but also minimize inevitable errors during manual annotation. The proposed method is here considered as an aide means to supervised learning rather than an alternative branch, where strong candidates for motion-induced anomalies can be provided in advance of human annotation. 

We found interesting observations that can be viewed as empirical evidence of the capability to reduce inevitable errors in the manual annotation for the motion influence of CVS. Figure \ref{pltool} (a) shows cases in which the proposed method successfully estimated abnormalities (blue dots), whereas some mistakes were present in the human-craft labels (non-red regions with blue dots). 

For quantitative examination, the following pilot study was conducted: A ten-year biosignal expert (Lee) was asked to reannotate one subject's CVS data with knowledge of abnormality predictions using the proposed method. Thereafter, we qualitatively measured the dissimilarity between the original and machine-guided annotations. There was inter-observer variability between them, and the machine-guided annotation approached the machine prediction, as shown in Figure \ref{pltool} (b) and (c). This observation somewhat demonstrates the capability of the proposed method to reduce human errors during manual annotation as well as minimizing cumbersome and time-consuming processes.

\section{Conclusion and Discussion}
In this study, we propose a novel ML-based CVS quality assessment method for a real-time hemodynamic monitoring system using MEI measurements, during which deliberate or inevitable motions cause significant loss of functionality for the extracted biophysical quantity. Motivated by minimizing the labor, time, and economic burdens associated with manual annotation, the proposed method is an unsupervised learning approach that provides a competitive alternative to supervised learning or, at least, an auxiliary means of labeling support. Its significance can be emphasized in the industrial field, where it is unavoidable to gather and utilize a large amount of CVS data to achieve high accuracy and robustness in real-world applications. The other core is to incorporate the heuristic perception of bioimpedance professionals into the model architecture, where the time context is the key to realizing motion influence in CVS. Two models, point-to-point and cycle-to-cycle, were presented, which were designed to have an explicit mechanism to reflect the time context of CVS data. Recognizing a group of CVSs during a cardiac cycle as a sequence point, the cycle-to-cycle model is advantageous to capture a significant heartbeat-related context and, thus, shows better performance.

The cycle-to-cycle model requires the use of complementary information from ECG. Fortunately, hemodynamic monitoring systems typically acquire ECG signals simultaneously because of their significance as vital signs. Thus, the use of ECG is practically reasonable.

In practice, one strategy to utilize the developed method is as follows: when a large amount of label-absent CVS data is provided, the method is applied to obtain pseudo-labels. The labeler subsequently annotates the dataset with reduced workflows, possibly to minimize human errors. The cut--off value $\tau$ in \eqref{assessment} is initially selected using the two-signal rule of thumb and can thereafter be adjusted at the labeler's discretion. If a small amount of paired data is available, $\tau$ may be determined by maximizing the $J$ statistic over the given small paired data. 

By utilizing the proposed method as a pseudo-labeling tool, it may be favorable to increase the TNR (small false positives) and even sacrifice the TPR (large false negatives). The main reason for this is to reduce missing true negatives when restricting manual annotation to parts around abnormalities estimated by the ML algorithm. Although the $J$ statistic is a good option for determining $\tau$, it may not be optimal. A solid strategy to optimize $\tau$ in terms of pseudo-labeling is an open question yet to be developed in our future studies.

\section*{CRediT authorship contribution statement}
C.M. Hyun: conceptualization, formal analysis, investigation, methodology, visualization, software, supervision, writing (original draft and review). T.-G. Kim: investigation, methodology, software, validation, visualization, and writing (Appendix B). K. Lee: data curation, formal analysis, validation, writing (review), and funding.

\section*{Declaration of Competing Interest}
The authors declare that they have no known competing financial interests or personal relationships that could have appeared to influence the work reported in this paper.

\section*{Data availability}
The data that support the findings of this study are available from one of the corresponding authors (K. Lee) upon reasonable request.

\section*{Acknowledgments}
We sincerely express our deep gratitude to BiLab company (Seongnam, Republic of Korea) for their help and collaboration. This work was supported by the Ministry of Trade, Industry and Energy (MOTIE) in Korea through the Industrial Strategic Technology Development Program under Grant 20006024.

\appendix
\section{Motion Artifact in Cardiac Volume Signal} \label{appen1}
\begin{figure}[h]
	\includegraphics[width=0.475\textwidth]{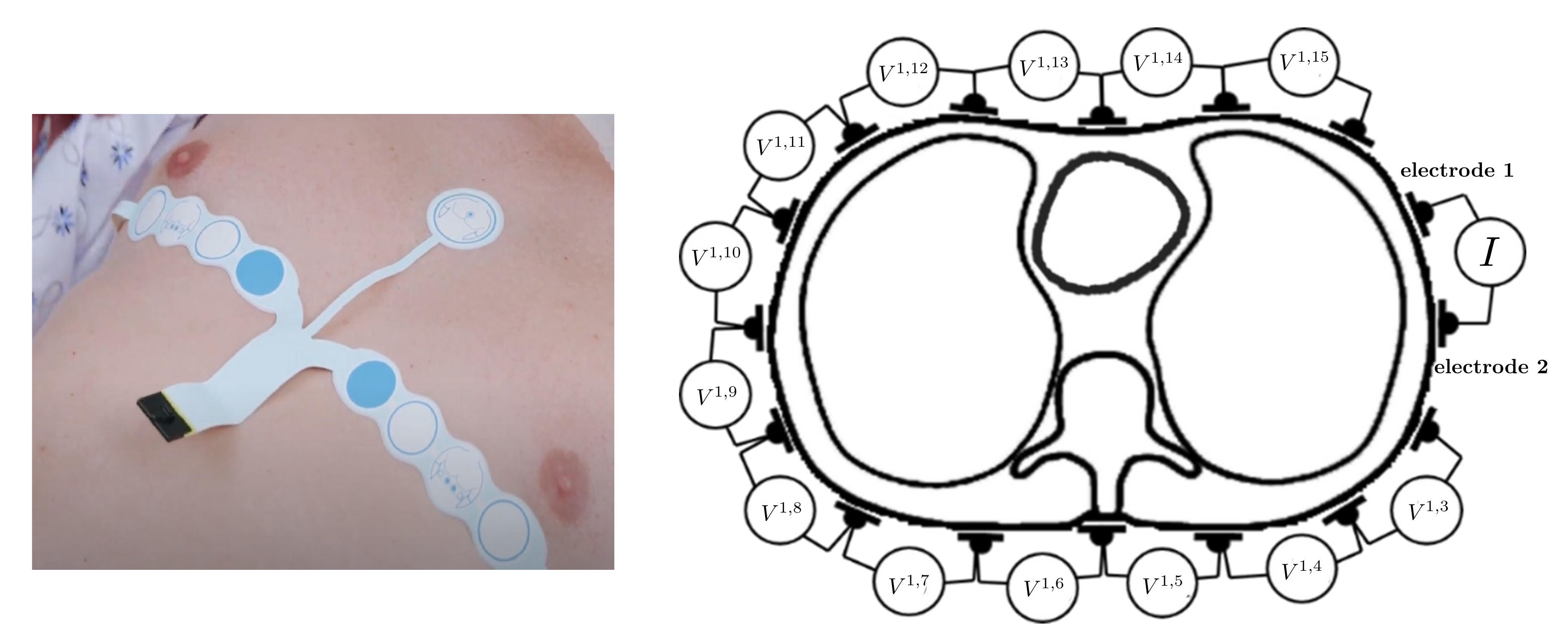}
	\caption{16-channel thoracic electrical impedance system for hemodynamic monitoring.}
	\label{appendix1}
\end{figure}

The hemodynamic monitoring system used in this study (Hemovista, BiLab, Republic of Korea) is a 16-channel thoracic electrical impedance device, which measures voltages using 16 electrodes attached around the human chest. See Figure \ref{appendix1}. Once an alternative current of $I$mA is injected from $i$-th to $(i+1)$-th electrodes, let $V^{i,j}_t$ be a voltage response between $j$-th and $(j+1)$-th electrodes. At sampling time $t$, the following trans-conductance $\boldsymbol G_t$ is obtained:
\begin{equation}\label{conductance} 
	\boldsymbol G_t = I \begin{bmatrix}  G^{1,3}_t, \cdots, G^{1,15}_t, \cdots, G^{16,2}_t,\cdots, G^{16,14}_t \end{bmatrix}^T,
\end{equation}
where $G^{i,j}_t$ is the reciprocal number of a real part of $V^{i,j}_t$ and $T$ represents the transpose operator. From $\boldsymbol G_t$, a CVS value $\boldsymbol x_t$ is obtained as follows. For some reference time $t_0$,
\begin{equation} \label{weighting}
\boldsymbol x_t = \boldsymbol w^{T} \dot{\boldsymbol G}_t \mbox{ and } \dot{\boldsymbol G}_t=\boldsymbol G_t - \boldsymbol G_{t_0},
\end{equation}
where the weighting vector $\boldsymbol w$ is designed to extract a cardiogenic component from $\boldsymbol G_t$ that is comprehensively affected by multiple sources including lungs and heart. Kindly refer to \cite{lee2021} for more details.

Based on the complete electrode model and Reynolds transport theorem, the trans-conductance $\boldsymbol G_t$ can be decomposed as follows \cite{hyun2023,lee2017}:
\begin{equation} \label{conduc_relation}
	\dot{\boldsymbol G}_t \approx \dot{\boldsymbol G}_{t}^{\mbox{\scriptsize normal}} + \dot{\boldsymbol G}_{t}^{\mbox{\scriptsize motion}},
\end{equation}
where 
\begin{align}
	|\dot{G}_{t}^{\mbox{\scriptsize normal},i,j}| & \propto \int_{\Omega} \dot{\gamma}_t(\boldsymbol \xi) \nabla u^i_t(\boldsymbol \xi) \cdot \nabla u_t^j(\boldsymbol \xi) d\boldsymbol \xi \label{normal} \\
	|\dot{G}_{t}^{\mbox{\scriptsize motion},i,j}| & \propto \int_{\partial \Omega} v_n(\boldsymbol \xi,t) \gamma_t(\boldsymbol \xi) \nabla u^j_t(\boldsymbol \xi)\cdot \nabla u^{k}_t(\boldsymbol \xi) d\boldsymbol s \label{motion}.
\end{align}
Here, $\Omega \subset \mathbb{R}^{3}$ is a time-varying human chest domain, \cor{}$v_n$ is an outward-normal directional velocity of $\partial \Omega$, $d\boldsymbol s$ is a surface measure, and $u_t^{i}$ and $\gamma_t$ are electric potential and conductivity distributions in $\Omega$, respectively. The relations \eqref{weighting} and \eqref{conduc_relation} yield
\begin{equation}\label{motion_rel}
	\boldsymbol x_t \approx \boldsymbol x_t^{\mbox{\scriptsize normal}} + \boldsymbol x_t^{\mbox{\scriptsize motion}},
\end{equation}
where $\boldsymbol x_t^{\mbox{\scriptsize normal}}= \boldsymbol w^{T}\dot{\boldsymbol G}_{t}^{\mbox{\scriptsize normal}}$ and $\boldsymbol x_t^{\mbox{\scriptsize motion}}=\boldsymbol w^{T}\dot{\boldsymbol G}_{t}^{\mbox{\scriptsize motion}}$. In the case of motion absence, $\boldsymbol x_t =\boldsymbol x_t^{\mbox{\scriptsize normal}}$ follows from $v_n=0$ in \eqref{motion}. In the presence of a large motion (i.e., $|v_n|$ grows), motion artifacts in CVS ($\boldsymbol x_t^{\mbox{\scriptsize motion}}$) become significant.

\section{Network and Training Details} \label{appen2}
This appendix provides details for network architectures and training procedures.

\paragraph{LSTM}
A LSTM model with a stacked length of $L$ can be represented as follows: For $k=1,\cdots,L$ and $\mathfrak T=t,\cdots,t+r$, 
\begin{align} \label{LSTM}
 	F_\mathfrak{T}^k & = \mbox{sig}(W_F^k \boldsymbol{x}_\mathfrak{T}^k + U_F^k C_{\mathfrak{T}-1}^k + b_F^k ) \nonumber \\
 	I_\mathfrak{T}^k & = \mbox{sig}(W_I^k \boldsymbol{x}_\mathfrak{T}^k + U_I^k C_{\mathfrak{T}-1}^k + b_I^k ) \nonumber \\
 	O_\mathfrak{T}^k & = \mbox{sig}(W_O^k \boldsymbol{x}_\mathfrak{T}^k +U_O^k C_{\mathfrak{T}-1}^k+b_O^k) \nonumber \\
 	C_\mathfrak{T}^k & = F_\mathfrak{T} \odot C_{\mathfrak{T}-1}^k + I_\mathfrak{T}^k \odot  \mbox{hyptan}(W_C^k \mathcal{\boldsymbol{x}}_\mathfrak{T}^k + b_C^k) \nonumber \\
 	H_\mathfrak{T}^k & = O_\mathfrak{T}^k \odot \mbox{hyptan}(C_\mathfrak{T}^k),
\end{align}
where $W$ and $U$ are weight matrices for a fully connected layer, $b$ is a bias vector, $\mbox{sig}$ is the sigmoid function, $\odot$ is the Hadamard product, and $\mbox{hyptan}$ is the hyperbolic tangent function. Here,  $\boldsymbol x^k_\mathfrak T$ is defined by
\begin{equation}
	\boldsymbol x^1_\mathfrak T = \boldsymbol x_\mathfrak T \mbox{ and } \boldsymbol x^{k}_\mathfrak T = H_{\mathfrak T}^{k-1} \mbox{ for } k \geq 2,
\end{equation}
where $\boldsymbol x_\mathfrak T$ is a point of an input sequence at time $\mathfrak T$.

\begin{figure}[t]
	\includegraphics[width=0.49\textwidth]{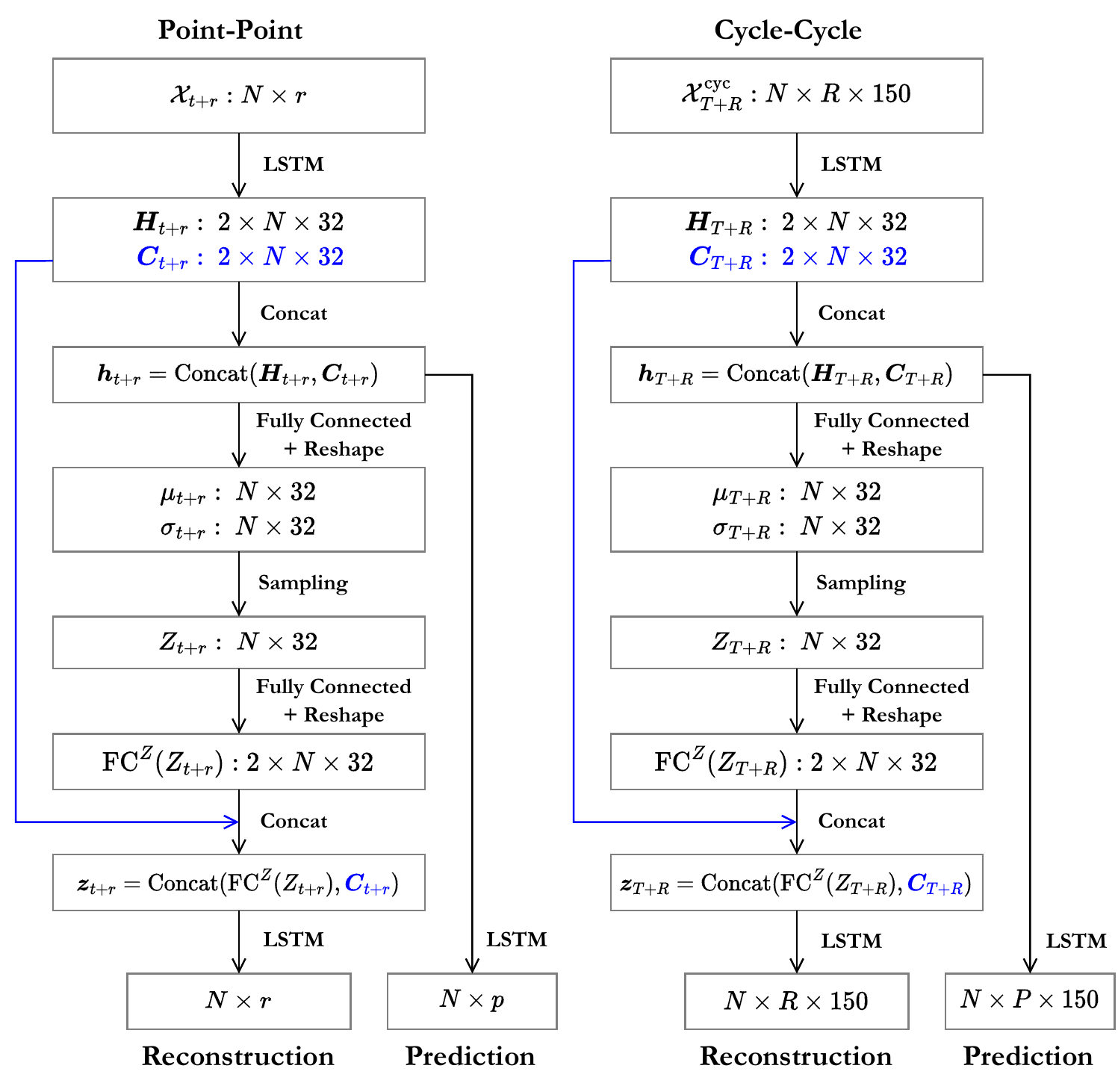}
	\caption{Network architecture details: (a) point-to-point and (b) sequence-to-sequence models. Here, $N$ is a symbol for the number of training sequence data.}
	\label{appendix2}
\end{figure}

\paragraph{VAE-LSTM Model}
In our implementation, we used LSTM with two stacked layers ($L=2$), each of which includes $32$ units. The latent dimension was assigned to be 32. The overall structure of VAE-LSTM is illustrated in Figure \ref{appendix2}. 

For an input sequence $\mathcal X_{t+r}$, the model is progressed as follows: A LSTM encoder $\mathcal E$ ingests a sequence of CVS data and subsequently outputs an encapsulated vector, which can be expressed as $\boldsymbol h_{t+r} = [\boldsymbol H_{t+r}, \boldsymbol C_{t+r}] \in \mathbb{R}^{32 \times 4}$, where
\begin{align}
	\boldsymbol H_{t+r}  = [H_{t+r}^{1}, H_{t+r}^{2}] \mbox{ and } \boldsymbol C_{t+r} = [C_{t+r}^{1}, C_{t+r}^{2}].
\end{align}
By duplicating $\boldsymbol h_{t+r}$, one copy is used for the reconstruction stage ($\mathcal D$) and the other for the prediction stage ($\mathcal D_{\mbox{\scriptsize pred}}$).

In the reconstruction stage, we input $\boldsymbol h_{t+r}$ to two fully connected layers with reshaping and then generate $\boldsymbol \mu_{t+r}$ and $\boldsymbol \sigma_{t+r}$. Afterwards, a latent vector $\boldsymbol Z_{t+r}$ is stochastically sampled in the sense of \eqref{sampling} through the reparametrization trick \cite{kingma2013} and passes through a fully connected layer with reshaping ($\mbox{FC}^{\boldsymbol Z}$), restoring the original data dimensionality. By employing $\mbox{FC}^{\boldsymbol Z}(\boldsymbol Z_{t+r})$ and $\boldsymbol C_{t+r}$ as initial hidden and cell states, the LSTM $\mathcal D$ generates a sequence, estimating the mirror of $\mathcal X_{t+r}$, from a dummy value ($\boldsymbol 0$), The output sequence can be expressed as 
\begin{equation} \label{decoder}
	\mathcal D(\boldsymbol z_{t+r}) = [H_t^{2},H_{t+1}^{2}, \cdots, H_{t+r}^{2}]
\end{equation}

For the prediction stage, in the same manner as in \eqref{decoder}, the LSTM decoder $\mathcal D_{\mbox{\scriptsize pred}}$ produces a sequence, predicting future $\mathscr X_{t+r,p}$, from a dummy value ($\boldsymbol 0$). Here, $\boldsymbol H_{t+r}$ and $\boldsymbol C_{t+r}$ in $\boldsymbol h_{t+r}$ are used as initial hidden and cell states.

\paragraph{Training Details}
For network training, the AdamW optimizer \cite{loshchilov2019} was consistently used, which is an extension of the Adam optimizer including a weight decay regularization to enhance precision of model parameter updates. We also used a learning rate scheduling strategy known as the one-cycle learning rate policy \cite{smith2018}. In our implementation, Weight \& Biases was utilized as a tool for logging the training process and optimizing hyperparameters. For the point-to-point models, we used a batch size of 1024, leaning rate of 0.01, weight decay rate of 0.01, and maximum epoch of 100. For the cycle-to-cycle models, we used a batch size of 128, leaning rate of 0.001, weight decay rate of 0.01, and maximum epoch of 100.

\printcredits
\bibliographystyle{cas-model2-names.bst}
\bibliography{reference.bbl}
\end{document}